\documentclass[pra,reprint,showpacs,showkeys,showpacs]{revtex4-1}
\usepackage{amsmath,amssymb,amsthm,times,graphics,graphicx,bm,subfigure}
\usepackage{amsmath,amssymb,amsfonts,bbm,graphicx,times}
\usepackage{amsthm}
\usepackage{amsfonts}
\usepackage{microtype}
\usepackage{mathrsfs}
\usepackage{placeins}
\usepackage{bbm}
\usepackage{booktabs}
\usepackage{color}
\usepackage[colorlinks,linkcolor=red,citecolor=blue,urlcolor=blue]{hyperref}

\newcommand{\be}{\begin{equation}}
\newcommand{\ee}{\end{equation}}
\newcommand{\ben}{\begin{eqnarray}}
\newcommand{\een}{\end{eqnarray}}
\newcommand{\bes}{\begin{subequations}}
\newcommand{\ees}{\end{subequations}}
\newcommand{\bF}{\begin{figure}}
\newcommand{\eF}{\end{figure}}

\def\ket#1{ \left| #1 \right\rangle}
\def\bra#1{{\langle #1 |  }}

\def\pd2v#1#2#3{\frac{\partial^2 #1}{\partial #2 \partial #3}}

\def\binom#1#2{\left( \begin{array}{c} #1 \\ #2 \end{array} \right)}

\def \2x2mat#1#2#3#4{
\left( \begin{array}{cc}
#1 &  #2 \\  #3 &  #4
\end{array} \right)
}

\def \i{i}

\begin{document}
\title{Qubit-Programmable Operations on Quantum Light Fields }

\author{Marco Barbieri$^{1}$, Nicol\`o Spagnolo$^2$, Franck Ferreyrol$^3$, R\'emi Blandino$^4$, Brian J. Smith$^5$, and Rosa Tualle-Brouri$^6$}
\affiliation{$^1$Dipartimento di Scienze, Universit\`a degli Studi Roma Tre, Rome, Italy }
\affiliation{$^2$Dipartimento di Fisica, Sapienza Universit\`a di Roma, Rome, Italy }
\affiliation{$^3$Laboratoire Photonique, Num\'erique et Nanostructures, Institut d'Optique, Talence, France}
\affiliation{$^4$School of Mathematics and Physics, the University of Queensland, St Lucia, QLD, Australia}
\affiliation{$^5$Clarendon Laboratory, Department of Physics, University of Oxford, United Kingdom}
\affiliation{$^6$Laboratoire Charles Fabry, Institut d'Optique, CNRS and Universit\'e Paris-Sud, Palaiseau, France and Institut Universitaire de France, Paris, France}

\begin{abstract}
{Engineering quantum operations is one of the main abilities we need for developing quantum technologies and designing new fundamental tests. Here we propose a scheme for realising a controlled operation acting on a travelling quantum field, whose functioning is determined by an input qubit. This study introduces new concepts and methods in the interface of continuous- and discrete-variable quantum optical systems.}
\end{abstract}

\maketitle
Control of quantum systems is a key task for any implementation of quantum technologies, as well as for fundamental tests~\cite{Wiseman}. Quantum optics, despite the fact that non-linear interactions are extremely hard to achieve for quantum light, has made considerable progress thanks to the adoption of measurement-induced nonlinearities~\cite{Knill}. This technique for inducing nonlinear behaviours in linear systems consists in utilising ancillary resources, and then perform conditional operations based on the outcome of a measurement on a part of the whole system. It has been mostly applied to the implementation of quantum gates for qubits encoded in degrees of freedom of single photons~\cite{Pittman,Obrien,Gasparoni,Bao,Lanyon,Shadbolt}.

Recent developments have shown that the same idea can be exploited for more elaborate control, when it is the state of the quantum field itself that is manipulated through measurement and post-selection~\cite{Ourjoumtsev,Erwan} . This is the case, for instance, of the operation of quantum gates in coherent-state quantum computing~\cite{NNiels,Anderson,Blandino}, entanglement distillation~\cite{NNiels1,Lvovsky}, photon addition and subtraction~\cite{Zavatta,Parigi,Kumar}, and noiseless amplification~\cite{Xiang,Ferreyrol,Zavatta1,Anderson1}. A new hybrid approach has built up from these investigations that aims at merging the advantages of a continuous-variable approach to quantum optics, with experimental and conceptual tools proper to single-photon manipulation, and viceversa~\cite{teleport,NNiels2,Jeong,Morin,Donati}.  

In this paper, we propose a different kind of interface between these two approaches in the concept of  qubit-programmed operation on a quantum field. Our proposal extends current methods for implementing photon addition and subtraction in order to operate with an arbitrary superposition, whose parameters can be set conditionally on the logic state of a qubit, encoded in a degree of freedom of a single photon. In principle, our  scheme can be embedded in a larger architecture, in which information processing is carried out on discrete variables, and eventually transferred to the quantum field. This adds new capabilities to the state engineering toolbox for quantum technologies.

\begin{figure}[b]
\includegraphics[width=\columnwidth]{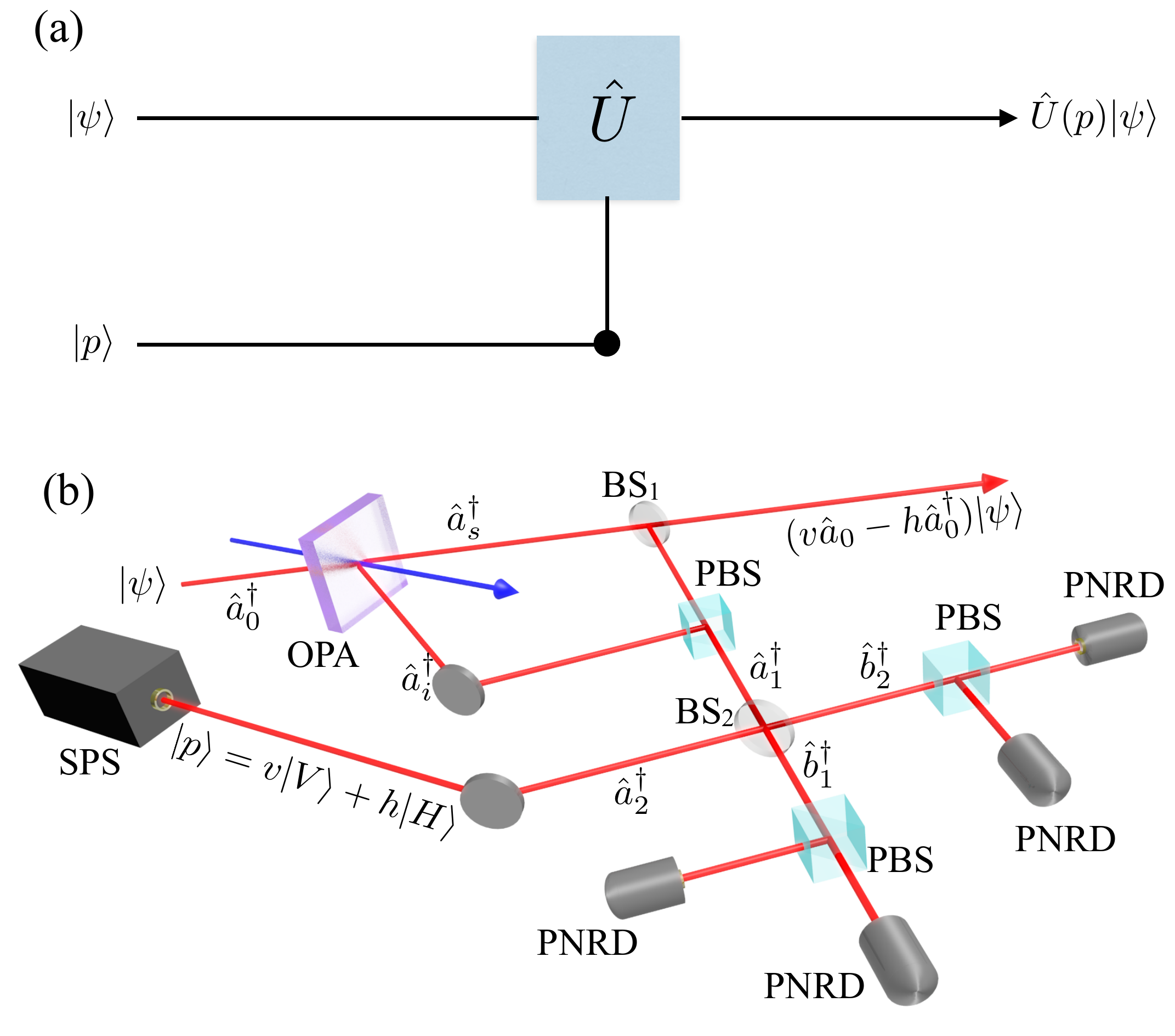}
\caption{a) Concept of the programmable gate; b) Scheme of the programmable device, indicating the labels of the different modes. BS$_1$ has transmissivity $t{\sim}1$. Output modes of the OPA and the single-photon source (SPS) are conveyed on the symmetric beamsplitter BS$_2$. Polarisation-resolved detection is performed by means of photon-number resolving detectors (PNRDs).}
\label{fig:setup}
\end{figure}

The general idea is illustrated in Fig.~\ref{fig:setup}a: a quantum field, described by a state $\ket{\psi}$, is the target we aim at manipulating trough an operation $\hat U(p)$, set according to the instructions $p$ we received from a second party; in other words, these instructions represent a programme that configures a particular choice of $\hat U(p)$. In the most general case, $\hat U(p)$ will be represented as a coherent superposition of some elementary operations. Therefore, the programme will need to be in the form of a quantum state $\ket{p}$, so that a proper mapping can be applied. Differently from conditional gates, such as the C-Not gate for two-qubits, we do not require that the programme state is preserved. Analogic control has been implemented using an optical displacement to modulate the probability amplitude of single-photon subtraction~\cite{NNiels,Anderson,coelho}; here we analyse a scheme for implementing the interface between a quantum optical field and the simplest discrete programme, a qubit~\cite{fiura}.

\begin{figure*}[t]
\centering
\includegraphics[viewport=0 0 1070 330, clip, width=2\columnwidth]{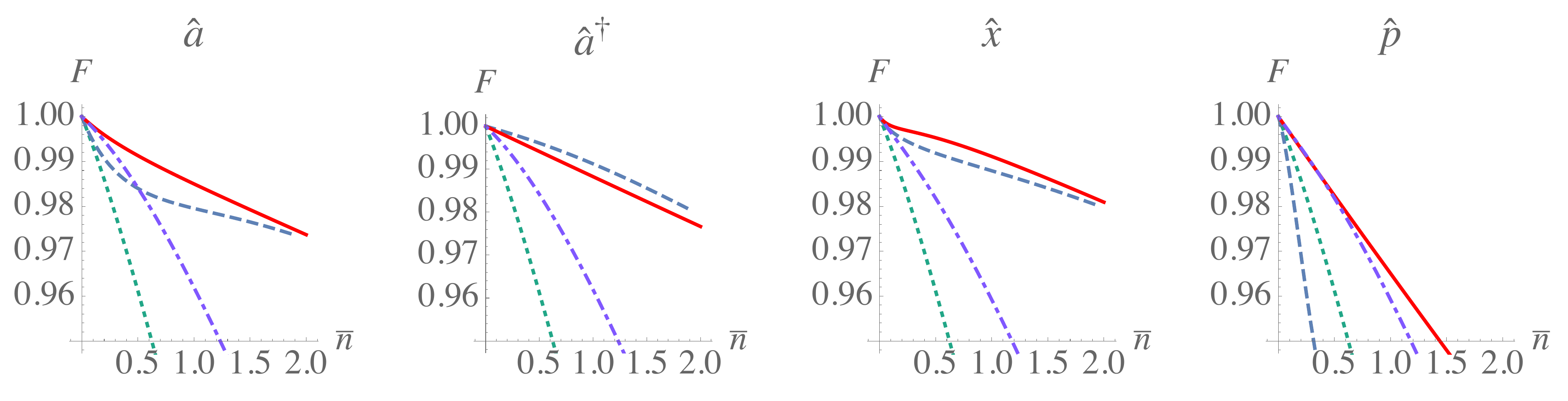}
\caption{Fidelity $F$ of the output states as a function of the average photon number $\bar n$ for four choices of the operator $\hat U$, and different families of input states: coherent states $\ket{\alpha}$ with $\alpha$ real solid red), cat states $\ket{\alpha}+\ket{-\alpha}$ (dashed blue), single-mode squeezed vacuum (dotted green). We also include the case when $U$ acts on a half of a two-mode squeezed state (dot-dashed purple). The transmissivity of the subtraction beamsplitter is $t{=}0.95$.}
\label{fig:fidelities}
\end{figure*}

At the basis of our proposal, there is the possibility of realising photon subtraction by means of a simple high transmissivity beamsplitter, and photon addition by using an optical parametric amplifier (OPA) in the low-gain regime. These operations can only be implemented probabilistically, with a single photon acting as a herald: in the former case, this comes from the reflected mode, in the latter, from the idler mode of the OPA. It is also known that these events can made indistinguishable by quantum interference, thus erasing the information as to whether the trigger signal was originated by an addition or a subtraction herald~\cite{Parigi1,LeeNha}. Inspired by the setup for quantum teleportation, we can use a single photon to control to what degree the erasure of information occurs.

Fig.~\ref{fig:setup}b details the apparatus of our proposal: consider the spatial mode $0$ prepared in the $n$-photon Fock state ${\ket{\psi}_0=\ket{n}_0}$, which constitutes the input to an OPA, driven in such conditions that approximate well photon addition. In terms of the two-mode interaction  $e^{g(\hat a_0^\dag \hat a_1^\dag+\hat a_0 \hat a_1)}$, this implies working in the regime of low parametric gain $g$. Further to the action of the OPA, the state is then passed through a high-transmissivity mirror ($t^2{\sim}1$, $r^2{\ll}1$) that realises photon subtraction. The two heralding modes are superposed on the spatial mode $1$, using two orthogonal polarisations: horizontal (H) for the subtraction herald, vertical (V) for the addition. The action of this device on $\ket{n}_0$ gives the following expression for the output state:
\begin{equation}
\begin{aligned}
\frac{C^{-(n+1)}}{\sqrt{n!}}\sum_{j=0}^{\infty}\frac{\Gamma^j}{j!}\sum_{k=0}^{n+j}\binom{n+j}{k}t^{n+j-k}r^k\times \\
\times (\hat a_0^\dag)^{n+j-k}( \hat a_{H1}^\dag)^{k}(\hat a_{V1}^\dag)^{j}\ket{0}_0\ket{0}_{H1}\ket{0}_{V1},
\end{aligned}
\end{equation}
which has been obtained by invoking the canonical transformations associated to the beam splitter, as well as the disentangling theorem~\cite{Collett}. Here, we have used the notation $C=\cosh{g}$, $\Gamma=\tanh{g}$, with $g$ the squeezing parameter. 

\begin{figure}[b]
\centering
\includegraphics[viewport=0 0 1000 450, clip, width=1.04\columnwidth]{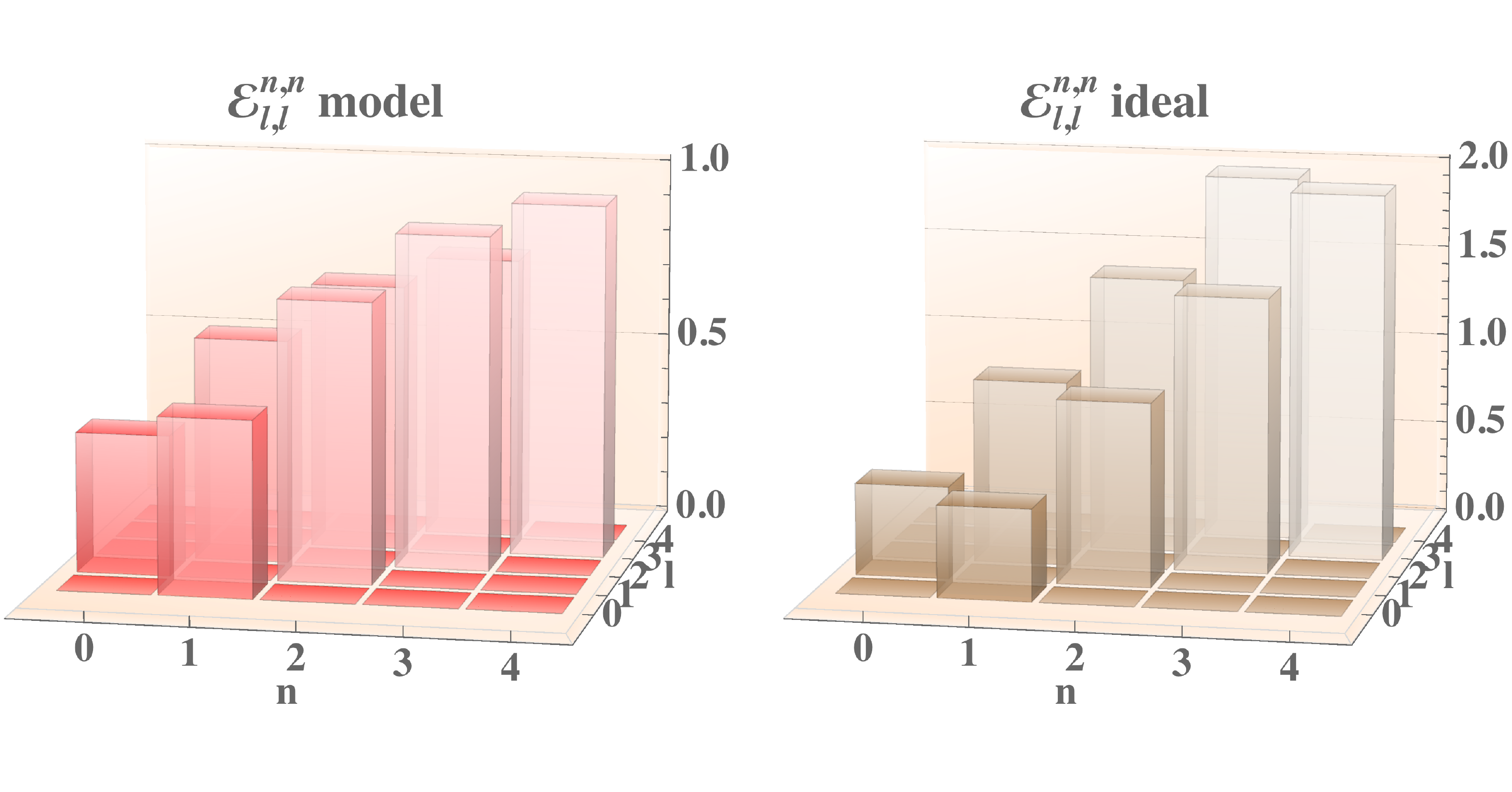}
\caption{Diagonal part of the process tensor ${\cal E}^{n,n}_{l,l}$ for both the modelled process Eq.\eqref{final} with $t=0.95$ (left), and the ideal $\hat x$ operation (right). Notice that, for ease of comparison, the modelled process tensor is rescaled by an overall factor 2/$r^2$.} 
\label{fig:process}
\end{figure}

Due to the correlations established between modes $0$ and $1$, any detection on the latter results in an operation applied to the input field; as an example, the detection of a single photon on mode $1$ on the polarisation $H$ ($V$) would herald the realisation of photon subtraction (addition). Detection in the diagonal polarisation basis would erase the information on the origin of the photon, thus the state on mode $0$ is transformed with the equal superposition $\hat a_0+\hat a_0^\dag$, and similarly for other choices of polarisation states, which can obtained by suitable choice of optical elements~\cite{Parigi1,LeeNha}. 

Our aim is to obtain the ability of controlling the choice of the superposition, programming it on a qubit in the form ${\ket{p}=h\ket{H}+v\ket{V}}$, {\it i.e} in the 'dual-rail' encoding: ${\ket{p}=\left(h\hat a^\dag_{H2}+v\hat a^\dag_{V2}\right)}\ket{0}_{H2}\ket{0}_{V2}$. To achieve our purpose, we superpose the two optical modes $1$ and $2$ on a beamsplitter with transmissivity $1/\sqrt{2}$, and then post-select on either the outcome $\ket{1}_{H1}\ket{0}_{V1}\ket{0}_{H2}\ket{1}_{V2}$ or $\ket{0}_{H1}\ket{1}_{V1}\ket{1}_{H2}\ket{0}_{V2}$. In an ordinary teleportation experiment, this would correspond to selecting the singlet in a Bell-state analysis. Consequently, the state of mode $0$ is projected to:
\begin{equation}
\frac{C^{-(n+1)}t^{n-1}}{\sqrt{2}}\left(v r\,\hat a_0-h \Gamma t^2\, \hat a^\dag_0\right)\ket{n}_0,
\label{final}
\end{equation}
therefore, in the limit $t\sim1$ and tuning the gain of the OPA so that $\Gamma= rt^{-2}$, we can map the state of the qubit $\ket{p}$ onto the operator $\hat U(p)=v\,\hat a_0-h\,\hat a_0^\dag$ which is acting on the input. Details on the calculations are presented in Apeendix A. Based on the transformation Eq.\eqref{final}, we can calculate the fidelity of the output states with the ideal case for different families of inputs; we report in Fig.~\ref{fig:fidelities} the results for four relevant cases $\hat a, \hat a^\dag, \hat x=\hat a+\hat a^\dag, \hat p=\i(\hat a-\hat a^\dag)$, for a transmission $t=0.95$, which is typical in such experiments~\cite{NNiels,Blandino,NonG,Ourjoumtsev1}.

\begin{figure}[t!]
\centering
\includegraphics[viewport=0 0 900 950, clip, width=1.04\columnwidth]{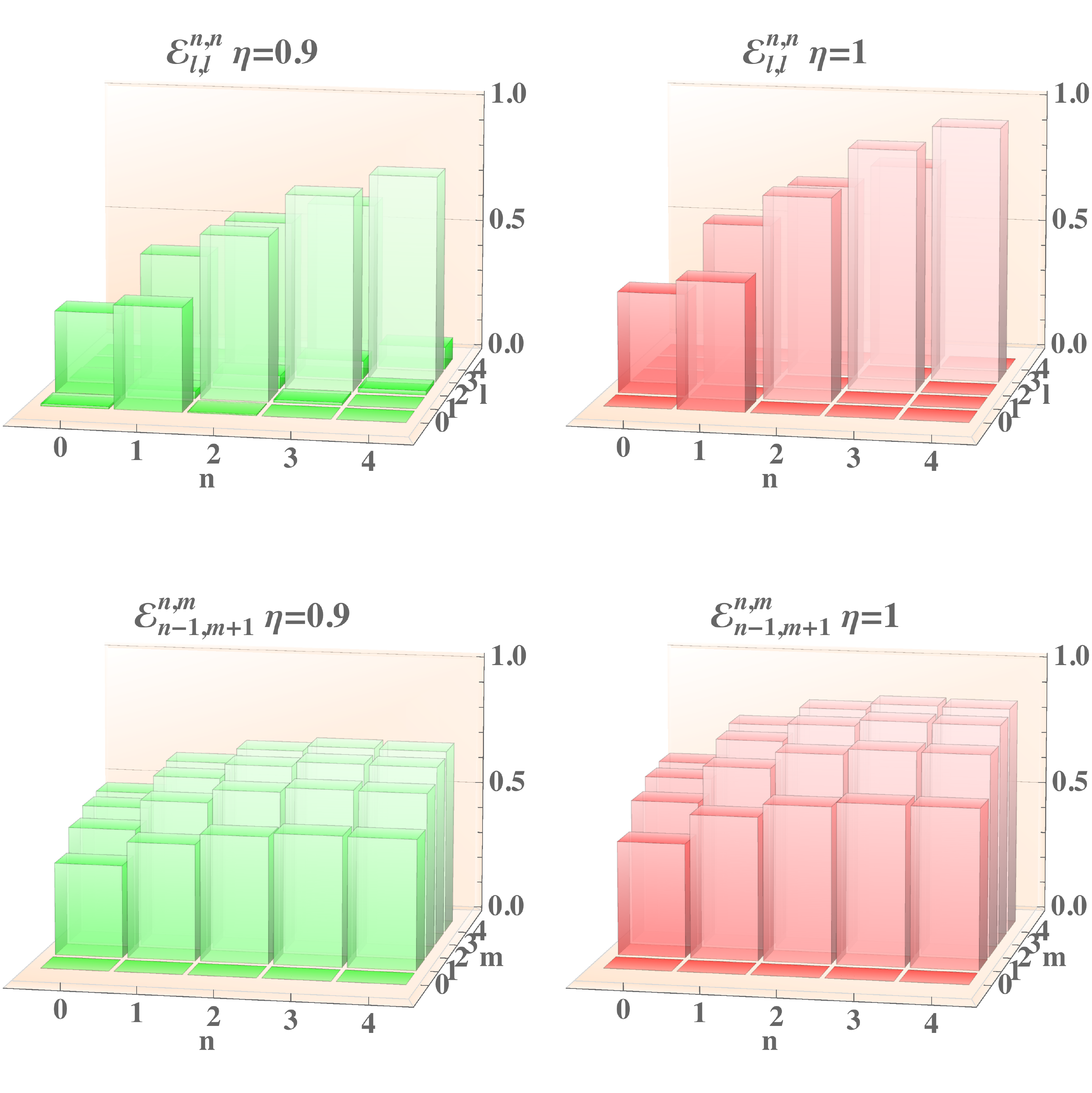}
\caption{Comparison of the process tensor at different values of quantum efficiency of the detectors $\eta$, which is assumed equal for all of them. Top: diagonal part of the process tensor ${\cal E}^{n,n}_{l,l}$. Bottom:  section representative of the coherence transfer ${\cal E}^{n,m}_{n-1,m+1}$. For ease of comparison, the modelled process tensor is rescaled by an overall factor 4/$r^2$.} 
\label{fig:processpiumonnezza}
\end{figure}

While a satisfactory level of fidelity can be reached at low photon numbers for different relevant classes of input states, the behaviour depends on the specific class. We can draw two general considerations: first, the quality of the approximation remains comparable to conditional operations commonly performed~\cite{NNiels,Anderson,Blandino,Parigi1}. Remarkably, the fidelity for coherent states depends on the relative phase between the state and the operator. We have indeed verified that $|\bra{\alpha}\hat x\, \hat U\left(\frac{\ket{H}{-}\ket{V}}{\sqrt{2}}\right)\ket{\alpha}|^2{=}|\bra{\i\alpha}\hat p\, \hat U\left(\frac{\ket{H}{+}\ket{V}}{\sqrt{2}}\right)\ket{\i\alpha}|^2$. Moreover, we observe that states presenting a long tail in the photon number distribution, such as single-mode squeezed states, are more affected by the departure of the conditioned process from the ideal one.

This occurrence can be understood by referring to the explicit form of the the process tensor ${\cal E}^{n,m}_{l,k}$, where each term is defined as the probability amplitude for $\ket{n}\bra{m}$ being transformed into $\ket{l}\bra{k}$~\cite{Lvovsky1}. In Fig.~\ref{fig:process}, our results for the process tensor associated to $\hat U\left(\frac{\ket{H}{-}\ket{V}}{\sqrt{2}}\right)$ are compared with the ideal case $\hat x$. When observing the diagonal terms ${\cal E}^{n,n}_{l,l}$, that govern the transfer of populations, one observes that the main departure is the attenuation factor $C^{-(n+1)}t^{n-1}$, which is also responsible for the unbalance between the addition and subtraction terms.

There are two main technical limitations which may hamper our scheme. The first is obtaining the correct mode matching between signal and idler modes of the OPA~\cite{Parigi1}; this can be achieved by properly choosing the nonlinear medium according to it dispersion properties, as well as an appropriate operating wavelength region~\cite{Peter,Brian,Christine,Justin}. The second is the limited detection efficiency of the heralding detectors; we have conducted a numerical analysis to include this effect into the expression of the process tensor ${\cal E}^{n,m}_{l,k}$, including the first-order corrections. We then imagine to operate in high-efficiency regime, as achieved in recent demonstrations~\cite{SaeWoo,Fiore,Ben}.

The results of this analysis are summarised in Fig.~\ref{fig:processpiumonnezza}, where we show a comparison of the process tensors for two values of the detection efficiency of the photon-number resolving detectors, $\eta{=}0.9$ and $\eta{=}1$ for the programme $\left(\ket{H}+\ket{V}\right)/\sqrt{2}$. Detail on the calculations are presented in Appendix B. For clarity of presentation, we only consider a single post-selection event on $\ket{1}_{H1}\ket{0}_{V1}\ket{0}_{H2}\ket{1}_{V2}$, with the other event giving qualitatively similar results.

\begin{figure}[t!]
\centering
\includegraphics[viewport=30 30 1100 560, clip, width=1.04\columnwidth]{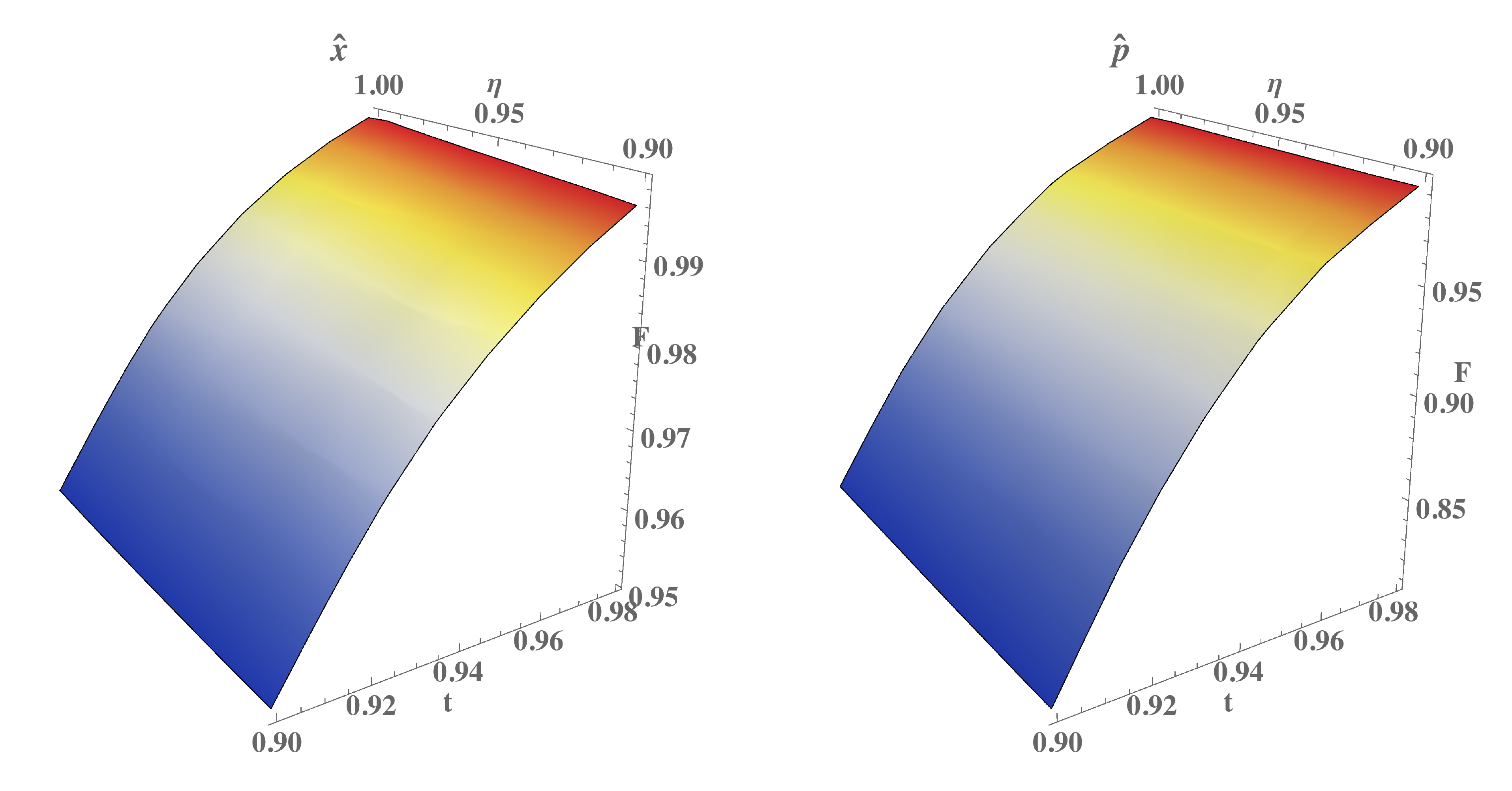}
\caption{Comparison of the fidelities between ideal and modelled outputs for coherent states inputs $\ket{\alpha}$ with $\alpha{=}1$ as a function of the transmissivity $t$ and the detection efficiency $\eta$ for $\hat U=\hat x$ and $\hat U=\hat p$.} 
\label{fig:rosa}
\end{figure}

The diagonal elements reveal that the attenuation at high photon number is more pronounced, since the inefficiency mainly results in the persistence of some population in original level $\ket{n}$; a minor effect consists in the transfer of population from $\ket{n}$ to $\ket{n\pm2}$, in analogy with the results in~\cite{a2b2}. Further insight is provided by observing the terms governing the coherence of the output state. The inefficiencies of the detection process induce a general attenuation of those terms, as one can not discriminate anymore among different incoherent detection events. The sheer effect is a reduction of the maximal photon number at which a satisfactory fidelity can be found with respect the results reported in Fig.~\ref{fig:fidelities}. This effect is captured more quantitatively in Fig.~\ref{fig:rosa}, in which we show the fidelities between ideal and modelled outputs for coherent state inputs at moderate average photon number $\bar n=1$. The protocol is then resistant to small detection loss, although it should be observe that these affect the overall success probability.

In conclusion, we have presented a scheme for controlling the quantum state of a travelling light field conditionally on the logical state of a qubit, which acts as a programme for a processing device. Our analysis highlights that a satisfactory level of fidelity can be achieved for relevant classes of states, in particular coherent states $\ket{\alpha}$ and cat states $\ket{\alpha}{+}\ket{-\alpha}$. We also investigate the effect on the process of imperfect detection, revealing how this limits the useful regime of photon number at the input. While high efficiencies are needed, steady progress of the detection technology~\cite{SaeWoo,Fiore,Ben} as well as quantum light sources~\cite{Justin,Harder} might allow the implementation of our protocol in the foreseeable future. This would allow for a more flexible control of quantum light, in particular for the investigation and employ of micro-macro quantum correlations~\cite{bruno,lvovsky12,kwon}.

{\it Acknowledgements} We thank M. Karpinski, A. Datta, M. Cooper, I.A. Walmsley, and F. Sciarrino for discussion and encouragement. M.B. is supported by a Rita Levi-Montalcini fellowship of MIUR. N.S. is supported by the ERC-Starting Grant 3D-QUEST (grant agreement no. 307783, http://www.3dquest.eu). BJS was partially supported by the Oxford Martin School programme on Bio-Inspired Quantum Technologies. We acknowledge support from FIRB Futuro in Ricerca HYTEQ, the EPSRC (Grants No. EP/K034480/1, No. EP/H03031X/1, and No. EP/H000178/1), the EC project SIQS, and the AFOSR EOARD.

\section{Appendix A}

We start our analysis of the qubit-programmable operation with the amplification in the Optical Parametric Amplifier (OPA). Its action is described by the operator:
\be
\label{17}
e^{g(\hat a^\dag_0 \hat a^\dag_1+\hat a_0 \hat a_1)} = e^{\Gamma \hat a^\dag_0 \hat a^\dag_1}C^{-(\hat n_0+\hat n_1+1)}e^{-\Gamma\hat a_0 \hat a_1},
\ee
where $\Gamma{=}\tanh g$, and $C=\cosh g$. By acting with the operator \eqref{17} on the input $\ket{n}_0\ket{0}_1$, we find:
\be
\label{18}
e^{g(\hat a^\dag_0 \hat a^\dag_1+\hat a_0 \hat a_1)} \ket{n}_0\ket{0}_2= e^{\Gamma \hat a^\dag_0 \hat a^\dag_1}C^{-(n+1)}\ket{n}_0\ket{0}_1.
\ee
We develop the exponential as a power series:
\be
\frac{C^{-(n+1)}}{\sqrt{n!}}\sum_{j=0}^{\infty}\frac{\Gamma^j}{j!}\left(\hat a^\dag_0\right)^{n+j}\left(\hat a^\dag_1\right)^j\ket{0}_0\ket{0}_1
\ee 
which describes the output from the OPA. Subtraction is implemented by means of a beam splitter (BS) with transmissivity $t\simeq1$, and reflectivity $r\leq1$, which couples the input mode 0 with a vacuum mode 2:

\be
\label{out}
\frac{C^{-(n+1)}}{\sqrt{n!}}\sum_{j=0}^{\infty}\frac{\Gamma^j}{j!}\sum_{k=0}^{n+j}t^{n+j-k}r^k\left(\hat a^\dag_0\right)^{n+j-k}\left(\hat a^\dag_1\right)^j\left(\hat a^\dag_2\right)^k\ket{0}_0\ket{0}_1\ket{0}_2
\ee 
Mode 1 and mode 2 are superposed on a single spatial mode, by mapping mode 2 into its horizontal ($H$) polarisation, and mode 1 into the vertical ($V$) polarisation. These modes are denoted as $H1$ and $V1$

We now add the qbit $h\ket{H}+v\ket{V}$, which is written in the dual-rail representation as:
\be
\label{4}
\left(h\hat a^\dag_{H2}+v\hat a^\dag_{V2}\right)\ket{0}_{H2}\ket{0}_{V2}
\ee
Spatial interference between mode 1 and the qubit mode is realised on a symmetric BS: 
\be
\begin{aligned}
\label{5}
\hat a^\dag_{H1} = \frac{\hat b^\dag_{H1}+\hat b^\dag_{H2}}{\sqrt{2}}\\
\hat a^\dag_{H2} = \frac{\hat b^\dag_{H2}-\hat b^\dag_{H1}}{\sqrt{2}}
\end{aligned}
\ee
and the transform for the $V$-polarisation are defined analogously. 

These formulae allow us to study the interference between the output state \eqref{out} with the qbit \eqref{4}; for the sake of simplicity, we isolate the term $\left(\hat a^\dag_{H1}\right)^k\left(\hat a^\dag_{v1}\right)^j\cdot \left(h \hat a^\dag_{H2}+v \hat a^\dag_{V2}\right)$, which is transformed by the action of the BS \eqref{5} as
\begin{widetext}
\be
\begin{aligned}
\left(\frac{1}{\sqrt{2}}\right)^{j+k}\sum_{x=0}^{j}\sum_{y=0}^{k} h\left(\left(\hat b^\dag_{H1}\right)^{y}\left(\hat b^\dag_{H2}\right)^{k-y+1}\left(\hat b^\dag_{V1}\right)^x\left(\hat b^\dag_{V2}\right)^{j-x}-\left(\hat b^\dag_{H1}\right)^{y+1}\left(\hat b^\dag_{H2}\right)^{k-y}\left(\hat b^\dag_{V1}\right)^x\left(\hat b^\dag_{V2}\right)^{j-x}\right)+\\
+v\left(\left(\hat b^\dag_{H1}\right)^{y}\left(\hat b^\dag_{H2}\right)^{k-y}\left(\hat b^\dag_{V1}\right)^x\left(\hat b^\dag_{V2}\right)^{j-x+1}-\left(\hat b^\dag_{H1}\right)^y\left(\hat b^\dag_{H2}\right)^{k-y}\left(\hat b^\dag_{V1}\right)^{x+1}\left(\hat b^\dag_{V2}\right)^{j-x}\right)
\end{aligned}
\ee
\end{widetext}

The indexes $k,y,j,x$ must be fixed in such a way to give a non-zero product with $\bra{1}_{H1}\bra{0}_{H2}\bra{0}_{V1}\bra{1}_{V2}$. For this, it is required that either $k{=}y{=}0, j{=}1, x{=}0$ or $j{=}x{=}0, k{=}1, y{=}1$; the two index set correspond to the second term (in ${-}h$) and to the third term (in $v$), respectively. If we now substitute these indexes in Eq.\eqref{out}, we obtain:
\be
\begin{aligned}
\label{22}
\frac{C^{-(n+1)}t^{n-1}}{2\sqrt{n!}} \left(vr n \left(\hat a_0^\dag\right)^{n-1}-h\Gamma t^2 \left(\hat a_0^\dag\right)^{n+1}\right)\ket{0}_0\\
=\frac{C^{-(n+1)}t^{n-1}}{2} \left(vr \hat a_0-h\Gamma t^2 \hat a_0^\dag\right)\ket{n}_0
\end{aligned}
\ee
The product with $\bra{0}_{H1}\bra{1}_{H2}\bra{1}_{V1}\bra{0}_{V2}$ delivers a similar expression.

\section{Appendix B}

We calculate the next-order correction when we introduce loss before the detectors. We assume that the efficiency $\eta$ is the same for all the devices, and that these retain their number-resolving ability in full. We isolate a single term in Eq.\eqref{21}, which we write in the general form 
\be
\label{24}
\left(\hat b^\dag_{H1}\right)^\alpha \left(\hat b^\dag_{H2}\right)^\beta \left(\hat b^\dag_{V1}\right)^\gamma \left(\hat b^\dag_{V2}\right)^\delta.
\ee
The effect of loss is modelled by a beam splitter of transmissivity $\sqrt{\eta}$, and we denote the loss modes as $\hat l^\dag_{H1}$, and so on. The expression is modified as
\be
\label{25}
\alpha\delta \eta (1-\eta)^{\frac{\alpha+\beta+\gamma+\delta-2}{2}}\hat b^\dag_{H1}\left(\hat l^\dag_{H1}\right)^{\alpha-1}\left(\hat l^\dag_{H2}\right)^{\beta}\left(\hat l^\dag_{V1}\right)^{\gamma} \hat b^\dag_{V2}\left(\hat l^\dag_{V2}\right)^{\delta-1}
\ee
where we selected the useful post-selection events with a single photon each on the $H1$ and $V2$ modes, and none on the $H2$ and the $V1$ modes. The partial trace on the loss modes lead to a pre-factor
\be
\label{26}
\alpha\delta \eta (1-\eta)^{\frac{\alpha+\beta+\gamma+\delta-2}{2}}\sqrt{(\alpha-1)!}\sqrt{\beta!}\sqrt{\gamma!}\sqrt{(\delta-1)!}
\ee
which takes into account the probability amplitude associated to each event.

We consider the events with one photon in excess with respect the desired number: $\ket{2}_{H1}\ket{0}_{H2}\ket{0}_{V1}\ket{1}_{V2}$, $\ket{1}_{H1}\ket{1}_{H2}\ket{0}_{V1}\ket{1}_{V2}$, $\ket{1}_{H1}\ket{0}_{H2}\ket{1}_{V1}\ket{1}_{V2}$, and $\ket{1}_{H1}\ket{0}_{H2}\ket{0}_{V1}\ket{2}_{V2}$ , {\it i.e.}, we insert in Eq.\eqref{25} and Eq.\eqref{26}, $\alpha{=}2, \beta{=}0, \gamma{=}0,\delta{=}1$, and so on, excluding the cases with no physical meaning for the indexes $k,y,j,x$. The results are summarised in the Table, where the sets of unphysical indexes are written in red.

These indexes deliver the corrections to the leading term \eqref{22} (which now appears with an overall probability amplitude $\eta$):
\begin{widetext}
\be
\text{for 2,0,0,1},\qquad \frac{\eta\sqrt{1-\eta}}{\sqrt{2}}C^{-(n+1)}t^{n-2} r\left(-h t^2 \Gamma (\hat n_0+1)+\frac{vr}{2}\hat a_0^2\right)\ket{n}_0,
\ee

\be
\text{for 1,1,0,1},\qquad \frac{\eta\sqrt{1-\eta}}{4\sqrt{2}}C^{-(n+1)}t^{n-2} r^2 v \,\hat a_0^2 \ket{n}_0,
\ee

\be
\text{for 1,0,1,1},\qquad -\frac{\eta\sqrt{1-\eta}}{4\sqrt{2}}C^{-(n+1)}t^{n+2} \Gamma^2 h \,\left(\hat a_0^\dag\right)^2 \ket{n}_0,
\ee

\be
\text{for 1,0,0,2},\qquad \frac{\eta\sqrt{1-\eta}}{\sqrt{2}}C^{-(n+1)}t^{n} \Gamma\left(vr (\hat n_0+1)-\frac{ht^2\Gamma}{2} \left(\hat a_0^\dag\right)^2\right)\ket{n}_0.
\ee

\newpage

\begin{tabular}{|p{0.15\textwidth}|p{0.2\textwidth}|p{0.2\textwidth}|p{0.2\textwidth}|p{0.2\textwidth}|}
\hline
($\alpha,\beta,\gamma,\delta$)= & (2,0,0,1)& (1,1,0,1) &(1,0,0,1)&(1,0,0,2)\\
\hline
\color{black} y=$\alpha$& {\color{red}y=2} &  \color{black}y=1 & {\color{red} y=1} & {\color{red} y=1}\\
\color{black} k-y+1=$\beta$& {\color{red} k=1} & \color{black}k=1 & {\color{red} k=0} & {\color{red} k=0}\\  
\color{black} x=$\gamma$& {\color{red} x=0} & \color{black}x=0 & {\color{red} x=1} & {\color{red} x=0}\\  
\color{black} j-x=$\delta$& {\color{red} j=1} &\color{black} j=1 & {\color{red} j=2} & {\color{red} j=2}\\  
\hline
\color{black} y+1=$\alpha$& {\color{black} y=1} & \color{black}y=0 & {\color{black} y=0} & {\color{black} y=0}\\
\color{black} k-y=$\beta$& {\color{black} k=1} & \color{black} k=1 & {\color{black} k=0} & {\color{black} k=0}\\  
\color{black} x=$\gamma$& {\color{black} x=0} & \color{black} x=0 & {\color{black} x=1} & {\color{black} x=0}\\  
\color{black} j-x=$\delta$& {\color{black} j=1} & \color{black} j=1 & {\color{black} j=2} & {\color{black} j=2}\\  
\hline
\color{black} y=$\alpha$& {\color{red}y=2} & \color{red}y=1 & {\color{black} y=1} & {\color{red} y=1}\\
\color{black} k-y=$\beta$& {\color{red} k=2} & \color{red} k=2 & {\color{black} k=1} & {\color{red} k=1}\\  
\color{black} x+1=$\gamma$& {\color{red} x=-1} & \color{red} x=-1 & {\color{black} x=0} & {\color{red} x=-1}\\  
\color{black} j-x=$\delta$& {\color{red} j=0} & \color{red} j=0 & {\color{black} j=1} & {\color{red} j=1}\\  
\hline
\color{black} y=$\alpha$& {\color{black}y=2} &\color{black} y=1 & {\color{black} y=1} & {\color{black} y=1}\\
\color{black} k-y=$\beta$& {\color{black} k=2} &\color{black} k=2 & {\color{black} k=1} & {\color{black} k=1}\\  
\color{black} x+1=$\gamma$& {\color{black} x=0} & \color{black} x=0 & {\color{black} x=1} & {\color{black} x=0}\\  
\color{black} j-x=$\delta$& {\color{black} j=0} &\color{black} j=0 & {\color{black} j=1} & {\color{black} j=1}\\  
\hline

\end{tabular}

\end{widetext}


\begin{thebibliography}{99}
\bibitem{Wiseman} H. M. Wiseman and G. J. Milburn, {\it Quantum Measurement and Control} (Cambridge University Press, Cambridge, 2009).
\bibitem{Knill} E. Knill, R. Laflamme, and G.J. Milburn, Nature (London) 409, 46 (2001).
\bibitem{Pittman} T. B. Pittman et al., Phys. Rev. A 68, 032316 (2003).
\bibitem{Obrien} J. L. O'Brien, G. J. Pryde, A. G. White, T. C. Ralph, and D.Branning, Nature (London) 426, 264 (2003).
\bibitem{Gasparoni} S. Gasparoni, J-W Pan, P. Walther, T. Rudolph, and A. Zeilinger, Phys. Rev. Lett, 93, 050204 (2004).
\bibitem{Bao} X.-H. Bao, T.-Y. Chen, Q. Zhang, J. Yang, H. Zhang, T. Yang, and J.-W. Pan. Phys. Rev. Lett. 98, 170502 (2007).
\bibitem{Lanyon} B.P. Lanyon et al., Nature Phys. 5, 134 (2009).
\bibitem{Shadbolt} P.J. Shadbolt et al., Nature Photon. 6, 45 (2011).

\bibitem{Ourjoumtsev} A. Ourjoumtsev, H. Jeong, R. Tualle-Brouri, and P. Grangier, Nature (London) 448, 784 (2007).
\bibitem{Erwan} E. Bimbard, N. Jain, A. MacRae and A. I. Lvovsky, Nature Photon. 4, 243 (2010).

\bibitem{NNiels} J.S.Neegaard-Nielsen et al., Phys. Rev. Lett. 105 053602 (2010).
\bibitem{Marek} P. Marek, and J. Fiur\'a\v sek, Phys. Rev. A. 82, 014304 (2010).
\bibitem{Anderson} A. Tipsmark, R. Dong, A. Laghaout, P. Marek, M. Je\v zek, and U.L. Andersen, Phys. Rev. A 84, 050301(R) (2011).
\bibitem{Blandino} R. Blandino, F. Ferreyrol, M. Barbieri. P. Grangier, and R. Tualle-Brouri, New J. Phys 14 013017 (2012).

\bibitem{NNiels1} H. Takahashi et al., Nature Photon. 4, 178 (2010).
\bibitem{Lvovsky} Y. Kurochkin, A.S. Prasad, and A.I. Lvovsky Phys. Rev. Lett. 112, 070402 (2014).


\bibitem{Zavatta} A. Zavatta, S. Viciani, and M. Bellini Science 306, 660 (2004).
\bibitem{Parigi} V. Parigi, A. Zavatta, M.S. Kim, and M. Bellini Science 317, 1890 (2007).
\bibitem{Kumar} R. Kumar, E. Barrios, C. Kupchak and A. I. Lvovsky, Phys. Rev. Lett. 110, 130403 (2013) 

\bibitem{Xiang} G. Y. Xiang, T. C. Ralph, A. D. Lund, N. Walk, and G. J. Pryde, Nature Photon.
\bibitem{Ferreyrol} F. Ferreyrol, M. Barbieri, R. Blandino, S. Fossier, R. Tualle-Brouri, and P. Grangier, Phys. Rev. Lett. 104, 123603 (2010).
\bibitem{Zavatta1} A. Zavatta, J. Fiur\'a\v sek, and M. Bellini, Nature Photon. 5, 52 (2011)
\bibitem{Anderson1} M.A. Usuga, et al., Nature Phys. 6, 767 (2010).
 
\bibitem{teleport} S. Takeda, T. Mizuta, M. Fuwa, P. van Loock, and A. Furusawa, Nature (London) 500, 315 (2013).
\bibitem{NNiels2} J.S. Neergaard-Nielsen, Y. Eto, C.-W. Lee, H. Jeong, H. and M. Sasaki, Nat. Photon. 7, 439-443 (2013).
\bibitem{Jeong} H. Jeong, et al., Nat. Photon. 8, 564 (2014).
\bibitem{Morin} O. Morin, et al., Nat. Photon. 8, 570 (2014).
\bibitem{Donati} G. Donati et al., Nat. Commun., in press (2014).

\bibitem{coelho} A.S. Coelho, L.S. Costanzo, A. Zavatta, C.Hughes, M. S. Kim, and M. Bellini, arXiv:1407:6644 (2014).

\bibitem{fiura} J. Fiu\v r\'asek, M. Du\v sek, and R. Filip, Phys. Rev. Lett. 89, 190401 (2002).

\bibitem{Parigi1} M.S. Kim, H. Jeong, A. Zavatta, V. Parigi, M. Bellini, Phys. Rev. Lett. 101, 260401 (2008); A. Zavatta, V. Parigi, M.S. Kim, H. Jeong, M. Bellini, Phys. Rev. Lett. 103, 140406 (2009).


\bibitem{LeeNha} S.-Y. Lee, and H. Nha, Phys. Rev. A. 82, 053812 (2010).

\bibitem{Collett} M.J. Collett, Phys.Rev. A. 38, 2233 (1988).

\bibitem{NonG} J.Wenger, R. Tualle-Brouri,and P. Grangier, Phys. Rev. Lett. 92, 153601 (2004).
\bibitem{Ourjoumtsev1} A. Ourjoumtsev, R. Tualle-Brouri, J. Laurat, and P. Grangier, Science 312, 83 (2006).

\bibitem{Lvovsky1} M. Lobino, D. Korystov, C. Kupchak, E. Figueroa, B. C. Sanders and A. I. Lvovsky, Science 322, 563 (2008); S. Rahimi-Keshari, A. Scherer, A. Mann, A. T. Rezakhani, A. I. Lvovsky and B. C. Sanders, New J. Phys. 13, 013006 (2011).

\bibitem{Peter} P.J. Mosley, J.S. Lundeen, B.J. Smith, P. Wasylczyk, A.B. U'Ren, C. Silberhorn, and I.A. Walmsley, Phys. Rev. Lett. 100, 133601 (2008).
\bibitem{Brian} O. Cohen, J.S. Lundeen, B.J. Smith, G. Puentes, P. Mosley, and Ian A. Walmsley, Phys. Rev. Lett. 102, 123603 (2009).
\bibitem{Christine} A. Eckstein, B. Brecht, and C. Silberhorn, Opt. Exp. 19, 13770 (2011); 
\bibitem{Justin} J.B. Spring, et al., Opt Exp. 21, 13522 (2013).

\bibitem{SaeWoo} F. Marsili, V. B. Verma, J. A. Stern, S. Harrington, A. E. Lita, T. Gerrits, I. Vayshenker, B. Baek, M. D. Shaw, R. P. Mirin, and S. W. Nam, Nature Photon. 7, 210 (2013).
\bibitem{Fiore}  A. Divochiy, et al. Nature Photon., 2, 302 (2008).
\bibitem{Ben}  B. Calkins et al., Opt. Exp. 21, 22657 (2013).


\bibitem{a2b2} S.Y. Lee, and H. Nha, Phys. Rev. A 85, 043816 (2012).

\bibitem{Harder} G. Harder, V. Ansari, B. Brecht, T. Dirmeier, C. Marquardt, and C. Silberhorn, Opt. Exp. 21, 13975 (2013).

\bibitem{bruno} N. Bruno, A. Martin, P. Sekatski, N. Sangouard, R.T. Thew, and N. Gisin, Nat. Phys. 9, 545 (2013)  
\bibitem{lvovsky12} A.I. Lvovsky, R. Ghobadi, Chandra, A. Prasad, and C. Simon, Nat. Phys. 9, 541 (2013).
\bibitem{kwon} H. Kwon, and H. Jeong arXiv:1410.6823 (2014).

\end{thebibliography}
\end{document}